# The investigation of the hydride superconductor's parabolic-like critical temperature under high pressure

P Tongkhonburi [1] P. Udomsamuthirun[1], A. Changjan[2] T Kruaehong[3]

[1]Prasarnmit Physics Research Unit, Department of Physics, Faculty of Science, Srinakharinwirot University, Bangkok 10110, Thailand.
[2] Department of Environmental Technology and Applied Science, Faculty of Science and Technology, Pathumwan Institute of Technology, Pathumwan, Bangkok, 10330, Thailand.
[3]. Department of Physics, Faculty of Science and Technology, Suratthani Rajabhat University, Suratthani, Thailand

**Abstract**

Under the weak coupling, we investigate the critical temperatures under pressure of $H_3S$, $LaH_{10}$, $CaH_6$, and $Tl_2Ba_2CaCu_2O_{8+\delta}$ superconductors. The superconducting mechanism takes into account the electron-phonon interaction as well as the Coulomb interaction. Under high pressure, the critical temperature equation is calculated as a function of the fractional volume of the unit cell, and the Birch-Murnaghan equation of state is used to determine the relationship between fraction volume and pressure. Using thsee equation, we can analyze the parabolic-like relationship between the critical temperature and pressure of a superconductor. The parabolic behavior of these superconductors' critical temperature versus pressure can fits well. The maximal critical temperature of $Tl_2Ba_2CaCu_2O_{8+\delta}$, $H_3S$, $LaH_{10}$, and $CaH_6$ superconductors are predicted to be 112 K at 7 GPa, 197 K at 140 GPa, 252 K at 143 GPa, and 207 K at 174 GPa, respectively.

## 1. Introduction

One of the most significant expectations of superconductors in current physics has been the existence of superconductors at ambient temperature. Since 1911, Onnes [1] has discovered superconductivity in mercury with a critical temperature of around 4.2 K, and in 1986, Bednorz and Muller [2] have discovered cuprate superconductor. After that, the physicist displayed the critical temperature, which is higher than liquid nitrogen's boiling point. In order to increase the critical temperature of the superconductor, one of the key variables that researchers intend to take into account is pressure. The $Tl_2Ba_2CaCu_2O_{8+d}$ ( Tl2212) superconductor experiment has demonstrated the effects of pressure on superconductivity with pressure up to 30 GPa, where the critical temperature displayed parabolic-like behavior with an elevated value of 114 K around 7 GPa [3]. At a pressure of 7 GPa, the maximum critical temperature in $YBa_2Cu_3O_{7-d}$ (Y123), was around 132 K. Over the whole range of oxygen concentration, the investigation was seen under pressures of up to 17 GPa, and the results showed parabolic-like dependences [4-5]. The $HgBa_2Ca_2Cu_3O_{8+d}$ (Hg1223) superconductor was found to have the greatest critical temperature, measuring 153 K at 22 GPa in a slightly underdoped sample [6] and 164 K at 31 GPa in an optimally doped sample.[7]. The extreme critical temperature found in a hydride superconductor has been shown at pressures greater than that of cuprate superconductors. The $H_3S$ superconductor has a critical temperature of 203 at 155GPa [8]. Calcium hydride ($CaH_6$) has critical temperatures of 220-235 K at 150 GPa [9] and 215 K at 172 GPa.[10]. The equation of state, which depicts a link between the volume of a unit cell and the pressure of a substance under pressure, is another experiment regarding the impact of pressure on physical characteristics. During measuring superconductivity in cuprate superconductors at high pressures, anisotropic behavior became a key factor. The bulk module, the compressibilities, the interrelationships of the crystal structure, and the anisotropy of the cuprate material are all still consistent with the Murnaghan equation of state [11]. In the $LaH_{10}$ superconductor, the equation of state and superconductivity at pressures up to 140 GPa were provided [12], and third-order Birch-Murnaghan fitting was used to account for pressure-volume data [13-14]. There have been reports on the $CaH_6$ superconductor's critical temperature dependency on pressure as well as its equation of state.[15]

According to the theoretical view on hydride superconductors, it was evident that electron-phonon interaction and Coulomb repulsion occurred when hydrogen-rich superconductors were under high pressure and in their superconducting state. The strong electron-phonon coupling and Coulomb potential were used for reporting on the $H_3S$ superconductor [16]. However, the isotope effect exponent was closer to the BCS framework according to the findings of an experiment using hydrogen and deuterium sulfide at high pressure [17]. The isotope effect exponent was noticeably seen close to the BCS model in the proposed $LaH_{10}$ superconductor [18] with critical temperature roughly 250 K at 170 GPa. There have been many suggestions to apply the weak-coupling model at high pressure with adjusted density of states and carrier dispersion relation [19-23] to explain the rise in critical temperature. The electron-phonon process is established as the essential framework for explaining superconductivity in the weak-coupling limit. Although a static electron-phonon interaction can be identified, the screening Coulomb interaction under high pressure caused by the electrical charge of the crystal structure can also collaborate. In conventional superconductor, the only important electron-phonon interaction is essential and a suitable approximation for the phonon spectrum, according to Morel and Anderson's model [24], which was developed after they researched the electron-electron interaction, including Coulomb repulsion. As a little decreasing mechanism for the critical temperature, Coulomb repulsion is employed. However, the hydride superconductors are subject to extremely high pressure. The Coulomb effect should be greater than before that the impact of Coulomb potential is taken into account.

Using the weak-coupling interaction model, we aim to explain the parabolic-like critical temperature of the cuprate and hydride superconductor in this investigation. Extending the BCS model with parameters under pressure enabled the calculation of the critical temperature formula. Using our derived formula and the Murnaghan equation of state, we compared the experimental data of the cuprate and the hydride superconductor with the findings. Finally, we demonstrated that our model could explain the critical temperature of both cuprate and hydride superconductors, which resembles a parabolic curve of critical temperature at high pressure.

**2. Model and calculation**

The BCS theory's weak-coupling framework, which is appropriate for our calculation, takes into account the impact of external pressure on the critical temperature of superconductors. We can derive the Green's function of the superconducting statec using the BCS Hamiltonian and the mean field theory as $G(k,\omega_n) = \dfrac{1}{i\omega_n - \varepsilon_k \tau_3 + \Delta_k \tau_1}$, where $\tau_1$ and $\tau_3$ are the Pauli matrices and $\omega_n$ is the Matsubara frequency. The gap equation, $\Delta_k = \sum_k V_{kk'} <C_{-k\downarrow}C_{k\uparrow}>$, is determined by the self-consistent equation and may be derived as

$$\Delta_k = -\sum_{k'} V_{kk'} \frac{\Delta_{k'}}{2\varepsilon_{k'}} \tanh(\frac{\varepsilon_{k'}}{2T}) \qquad (1)$$

Here, the carrier energy $\varepsilon_k$ is measured from the Fermi energy. $\Delta_k$ is the superconducting gap.

In our calculation, the multi-interaction model accounted for the Coulomb effect. The mechanisms of superconductors are the attractive electron-phonon interaction $V_{ph}$ and the repulsive Coulomb interaction $U_c$, with the distinct cutoff energies of Debye phonon ($\omega_D$) and Coulomb interaction ($\omega_c$), respectively. It is suggests that the muti-interaction potential model of carrier $V_{kk'}$ are [25,26]: $V_{kk'} = -V_{ph} + U_c$ for $0 < |\varepsilon_k| < \omega_D$, and $V_{kk'} = +U_c$ for $\omega_D < |\varepsilon_k| < \omega_c$. The superconducting order parameter should be written in the similar behaviour as $\Delta_k = \Delta_{ph}$ for $0 < |\varepsilon_k| < \omega_D$, And $\Delta_k = \Delta_c$ for $\omega_D < |\varepsilon_k| < \omega_c$.

To incorporate pressure into our model, we assume that pressure can affect superconductors in two distinct ways: either by altering the density of state or by disturbing the carrier's energy dispersion. The narrow fluctuation constant observed in the form of the delta function also appeared in the density of state under pressure as form [19-23]

$$N(\varepsilon) = N(0)(1 + \chi\delta(\varepsilon - \varepsilon_0))  \quad (2).$$

Here, $\chi$ is the height of this fluctuation function and the shifted position from the unpressured state is set as $\varepsilon_0$ below the Fermi level. The density of state can be reduced to the BCS scenario by setting $\chi = 0$. As has been determined that pressure has an impact on the carrier dispersion relation. Due to the size of the volume distorting the crystal structure, unit cells now contain additional energy from external pressure. Ref.[19–23] states that they can extend the new state in terms of external pressure ($p$). Expanding the new state in a power series of the fraction volume $v$ ($v = \dfrac{V}{V_0}$) is the most practical technique to connect to the Murnaghan equation of state [13,14].

Therefore, if the new stable state of the carrier's dispersion relation is $\varepsilon_k(p)$, this may be extended to become

$$\varepsilon_k(p) = \varepsilon_k(0) + p\left[\frac{d\varepsilon_k(p)}{dp}\right]_{p=0} + \frac{p^2}{2}\left[\frac{d^2\varepsilon_k(p)}{dp^2}\right]_{p=0} + \ldots .$$

The influence of additional pressure on volume is not a linear term, hence the power order of this relationship is assumed, in accordance with the Murnaghan equation of state [13,14]. We intend for $p \alpha \dfrac{1}{v^\beta}$ to come out in $\varepsilon_k(p) = \varepsilon_k(0) + k(\dfrac{1}{V^\beta} - \dfrac{1}{V_0^\beta})\left[\dfrac{d\varepsilon_k(p)}{dp}\right]_{p=0}$, here $Q_e = \dfrac{k}{V_0}\left[\dfrac{d\varepsilon_k(p)}{dp}\right]_{p=0}$, $v = \dfrac{V}{V_0}$, for $V$ near $V_0$, and $V_0$ is the volume at ambient pressure. Following the expansion's first term, we can write

$$\varepsilon_k(v) \approx \varepsilon_k(V_0) + Q_e(\frac{1}{v^\beta} - 1)  \quad (3).$$

We employ Eqs. (1-3) as our primary equations to determine the critical temperature. After matrix setup and gap equation secular equation solution, we obtain

$$I_{11} = \frac{1}{\lambda_{ph} - \mu_c^*}  \quad (4).$$

Where $I_{11} = I_{21} = \displaystyle\int_0^{\omega_D} d\varepsilon \, \frac{1}{\varepsilon + Q_e(\frac{1}{v^\beta}-1)} \tanh\left(\frac{\varepsilon + Q_e(\frac{1}{v^\beta}-1)}{2T_c}\right) + \frac{\chi}{\varepsilon_0 + Q_e(\frac{1}{v^\beta}-1)} \tanh\left(\frac{\varepsilon_0 + Q_e(\frac{1}{v^\beta}-1)}{2T_c}\right)$ (5).

And the pseudo Coulomb interaction potential is $\mu_c^* = \dfrac{-\mu_c}{1+\mu_c I_{22}}$.

We are able to get the formula for the critical temperature as

$$T_c = 1.13(\omega_D + Q_e(\tfrac{1}{v^\beta}-1))e^{\left(-\dfrac{1}{\lambda_{ph}-\mu^*}+\chi\dfrac{\tanh\left(\dfrac{\varepsilon_0+Q_e(\tfrac{1}{v^\beta}-1)}{2T_c}\right)}{\varepsilon_0+Q_e(\tfrac{1}{v^\beta}-1)}-\int_0^{\omega_D}\dfrac{\tanh\left(\dfrac{\varepsilon+Q_e(\tfrac{1}{v^\beta}-1)}{2T_c}\right)}{\varepsilon+Q_e(\tfrac{1}{v^\beta}-1)}d\varepsilon\right)} \quad (5)$$

We can estimate the term involved with $Q_v(\tfrac{1}{v^\beta}-1)$ in integration into two possible scenario for $\left|Q_e(\tfrac{1}{v^\beta}-1)\right| > 2T_c$ and $\left|Q_e(\tfrac{1}{v^\beta}-1)\right| < 2T_c$ that may provide the solution of the integration

$$\int_0^{\omega_D} d\varepsilon \dfrac{\tanh\left(\dfrac{\varepsilon+Q_v(\tfrac{1}{v^\beta}-1)}{2T_c}\right)}{\varepsilon+Q_v(\tfrac{1}{v^\beta}-1)} \text{ as } \ln\left(\dfrac{Q_v(\tfrac{1}{v^\beta}-1)}{2T_c}\right) \text{ and } \dfrac{Q_v(\tfrac{1}{v^\beta}-1)}{2T_c}, \text{ respectively.}$$

The equation for the critical temperature of a superconductor at high pressure is Eq. (5), which demonstrates the relationship between the critical temperature and the fraction volume. And we have a relationship between fraction volume and pressure in the Murnaghan equation of state, which may relate to the experimental data of superconductors under high pressure. In order to demonstrate a relationship between the critical temperature and external pressure, our calculation employs Eq.(5) and the Murnaghan equation of state.

3. **Result and discussion**

In order to understand the connection between the critical temperature and the external pressure, we then estimate the critical temperature of the cuprate superconductors and the hydride superconductors using equation (5) and the Murnaghan equation of state. The Birch-Murnaghan equation of state, one variant of the Murnaghan equation of state, the measurement on pressures to determine the volume, was the equation of state that was applied in our calculation. We use the Birch-Murnaghan [13,14] as

$$P(v) = \dfrac{3B_0}{2}[v^{-\tfrac{7}{3}} - v^{-\tfrac{5}{3}}]\{1 + \dfrac{3}{4}(B_0' - 4)[v^{-\tfrac{2}{3}} - 1]\} \quad (6)$$

Here, the volume fraction is define $v = \dfrac{V}{V_0}$ that $V_0, B_0$ and $B_0'$ are the equilibrium cell volume, the bulk modulus and the derivative of bulk modulus with respect to pressure.

The cuprate superconductor has been one of the most remarkable superconductors over the past ten years. Many physicists have an interest in the parabolic-like critical temperature versus pressure. We begin by applying our model to cuprate superconductors, $Tl_2Ba_2CaCu_2O_{8+d}$, whose unit cell volume changes slightly at high pressure and the volume fractions are nearly one which the volume varies in range 370-435 $A^3$, and whose data are established. For the $Tl_2Ba_2CaCu_2O_{8+d}$ superconductor, which has almost all the experimental data, we apply our model to explain this behavior. The bilayer single crystal of the superconductor $Tl_2Ba_2CaCu_2O_{8+d}$ has been studied at pressures up to 30 $GPa$ through investigation of the lattice parameter, unit-cell volume, and critical temperature with no structure change in cell parameters; consequently, no structure phase transition was discovered in this material [3]. The Eq.(5) and Eq.(6) are used for numerical calculation and comparison to experimental data [3,27,28]. In Figure 1., we have the Birch-Murnaghan with $B_0 = 111.7$, and $B_0' = 4$ [3]. and the parameters used are (solid line): $\chi = 460$ $\varepsilon_0 = 10$ $\lambda = 0.32$ $\mu = 0.01$ $\omega_D = 300$ $\omega_c = 350$ $Qe = -200$, $\beta = 3.3$. Our calculation can get the beautiful parabolic-like and perfect consistent with the experimental results. The maximum critical temperature is about 112 K at 7 $GPa$ that agree with Ref.[3] which the maximum is found of 114 K at 6.8 $GPa$.

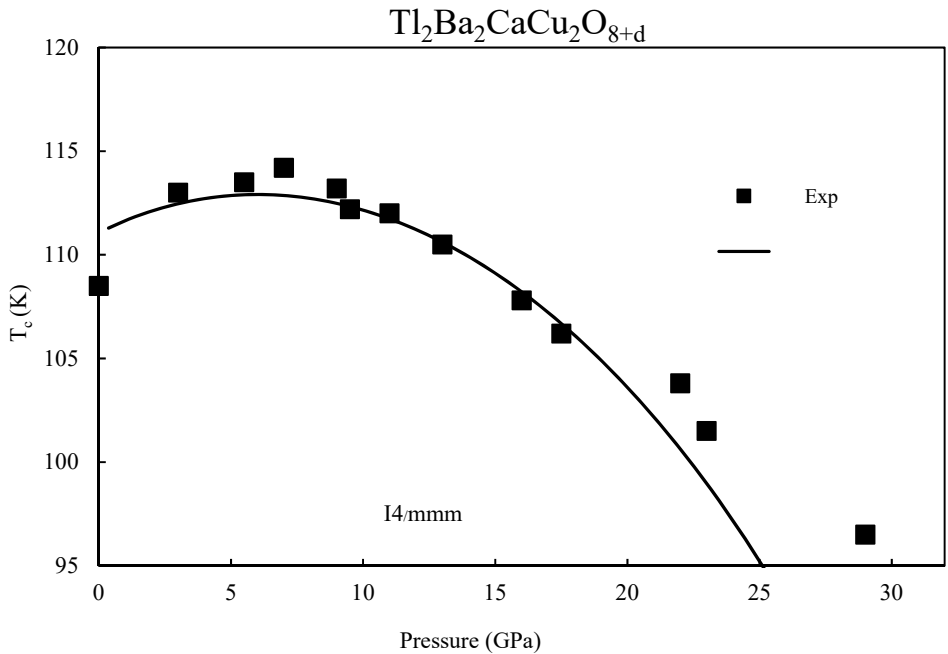

Figure .1 The critical temperature of $Tl_2Ba_2CaCu_2O_{8+d}$ superconductor was shown together with the calculation (solid line) and experimental data [3,27,28] (solid square).

The $H_3S$, LaH$_{10}$, and CaH$_6$ superconductors for hydride superconductors are of particular interest to us since they can exhibit the highest critical temperature with a parabolic-like form under high pressure. At ambient pressure, they are virtually in the gas phase, when they transition into the solid phase that superconductivity begins to appear. Since the fraction volumes of hydride superconductors are smaller than 1, we can modify these constraints by determining the appropriate $\beta$ value of the variable in our $p \alpha \frac{1}{v^{\beta}}$ assumption. The remaining pressure settings are being examined until the results of our calculations and the experiments agree. The Debye cutoff is obtained from each hydride's data, and the Coulomb cutoff is set to be greater than the Debye cutoff. The electron-phonon coupling constant was in the case of weak coupling. And, after doing several sampling calculations, we realized that the Coulomb coupling constant had little effect on our calculations, therefore just a modest quantity of Coulomb coupling constant was used.

The experimental results for the hydrogen hydride superconductor are given in Figure 2 as solid triangles and squares, and our calculations using Eq.(5) and Eq.(6) are shown as solid lines. This material contains a variety of crystal phase structures. We take particular attention to the two crystallographic phases in this material that change when the pressure rises from Cccm to Im-3m [8,29-32]. However, as there are insufficient data to definitively indicate the transition line, a mixed phase is postulated to exist between the two phase regimes. The Birch-Murnaghan $\beta_0 = 86.63$ and $\beta_0' = 3.9$ [32] were obtained by analyzing the lattice parameter, unit-cell volume, and critical temperature vs pressure up to 220 GPa. The $\chi = 600$, $\varepsilon_0 = 250$, $\lambda = 0.3$, $\mu = 0.01$, $\omega_D = 870$, $\omega_c = 970$, $Qe = -1.75$, $\beta = 4.0$, $\beta_0 = 10$, $\beta_0' = 4.2$ are the parameters used in Figure 2. . In phase of mixed and Im-3m phase, our calculation can display the parabolic-like and is highly compatible with the experimental data. The lower critical temperature in the Cccm phase prevents our calculation from fitting the data effectively. In the Cccm phase, we expect that the highest critical temperature will be around 197 K at 140 GPa.

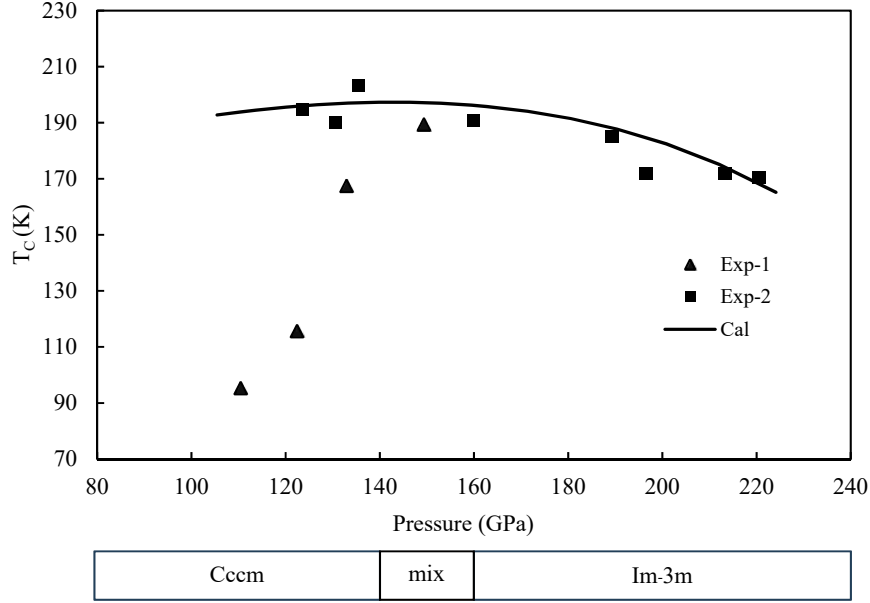

Figure.2. The calculation (solid line) and experimental data (solid triangles and squares) [8,29-32] regarding the critical temperature of a hydrogen hydride superconductor were shown.

The experimental LaH$_{10}$ data are compared with our calculation in Figure 3. As pressure rises, there exist three phases: C2/m, mixed, and Fm-3m phase. Experimental data are shown as solid dots [33-35] and the calculations as solid and dashed lines. The Fm-3m phase is a high-symmetry phase that is also found in areas of low pressure. Only the lower pressure area contains the lower phase C2/m. The lattice parameter, unit-cell volume, critical temperature vs pressure up to 220 GPa, were used to determine the Birch-Murnaghan, $B_0 = 27$ and $B'_0 = 4$ [12]. There are two lines calculated for LaH$_{10}$ superconductor: a solid line for greater pressure and a dashed line for lower pressure zones. The parameter used are solid line: $\chi = 420$ $\varepsilon_0 = 100$ $\lambda = 0.5$, $\mu = 0.01$, $\omega_D = 700$, $\omega_c = 800$, $Qe = -4.7$, $\beta = 4.0$, $\beta_0 = 57$, $\beta'_0 = 3.1$ and dashed line: $\chi = 520$, $\varepsilon_0 = 100$, $\lambda = 0.42$, $\mu = 0.01$, $\omega_D = 700$, $\omega_c = 800$, $Qe = -4.1$, $\beta = 4.2$, $\beta = 20$, $\beta'_0 = 4.0$. The calculations and experimental findings were quite consistent. Take into account that the calculation result in the Fm-3m phase region was parabolic-like and could forecast the highest critical temperature at around 252 K at 143 GPa, which is consistent with the experiment's findings of roughly 250 K at 150-170 GPa. While the calculation for the R3m phase appeared to show a linear relationship with the greater critical temperature expected to be higher than for the Fm-3m phase. The critical temperature under varying pressure can be effectively matched within the 2 sets that comprise our parameters.

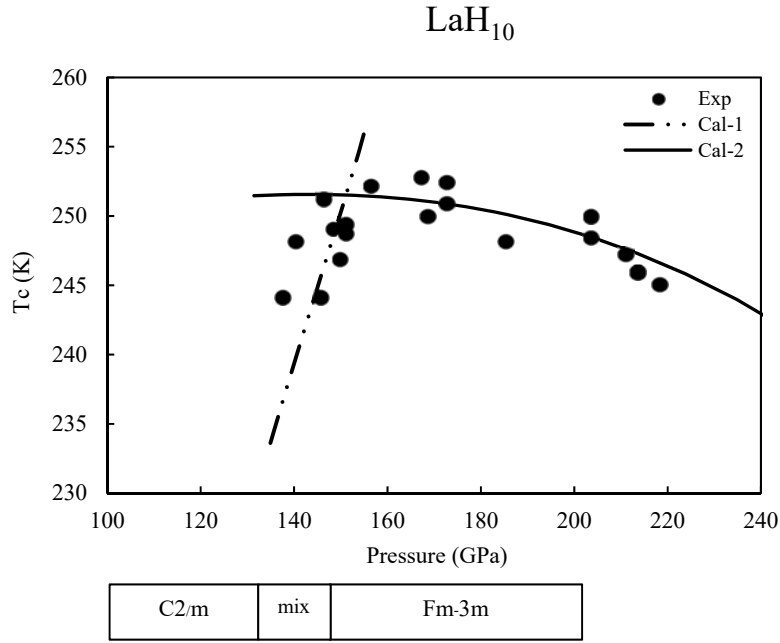

Figure .3 The calculation (solid and dashed lines) and experimental data (solid dots) [33-36] of critical temperature of LaH$_{10}$ superconductor were shown.

The CaH$_6$ superconductor's calculated and experimental data are displayed in Figure 4 along with the relationship between critical temperature and pressure. In this superconductor, there are two phases known as P2$_1$/m and Im-3m phases [9–10] that these phases are stable at the pressure 50-100 and 150-200 GPa, respectively. The lattice parameter, unit-cell volume, critical temperature versus pressure up to 220 GPa were done which the Birch-Murnaghan: $B_0 = 221$ and $B'_0 = 3$ [15]. The unit cell capacity changes from 24-20 A$^3$ as the pressure ranges from 110 to 220 GPa. Because of the anisotropic stress present, which causes varied distortion in various attempts, a broad range of critical temperatures were discovered to be between 100 and 220 $K$. Due to the large range of critical temperatures and the existence of two phase transitions, we divided our calculation into two portions for P2$_1$/m and Im-3m, respectively: a solid line and a dashed line. Following some manifestation, we can identify the optimal consistency for both the experimental and the computational parts. The parameter used are solid line: $\chi = 550$, $\varepsilon_0 = 23$, $\lambda = 0.33$, $\mu = 0.01$, $\omega_D = 960$, $\omega_c = 1060$, $Qe = -3.7$, $\beta = 3.80$, $\beta_0 = 97$, $\beta'_0 = 3.00$ and dash line: $\chi = 380$, $\varepsilon_0 = 23$, $\lambda = 0.375$, $\mu = 0.01$, $\omega_D = 960$, $\omega_c = 1060$, $Qe = -2.85$, $\beta = 3.97$, $\beta_0 = 130$, $\beta'_0 = 3.00$. The critical temperature in the P2$_1$/m Phase rises as pressure rises, which higher critical temperature should be found. Additionally, the critical temperature for the Im-3m phase is seem to be constant, with a little parabolic-like observation. In comparison to the experiment, which discovered a maximum critical temperature of 215 $K$ at 172 $GPa$ in the Im-3m phase, the maximum critical temperature may be expected to be about 207 $K$ at 174 $GPa$.

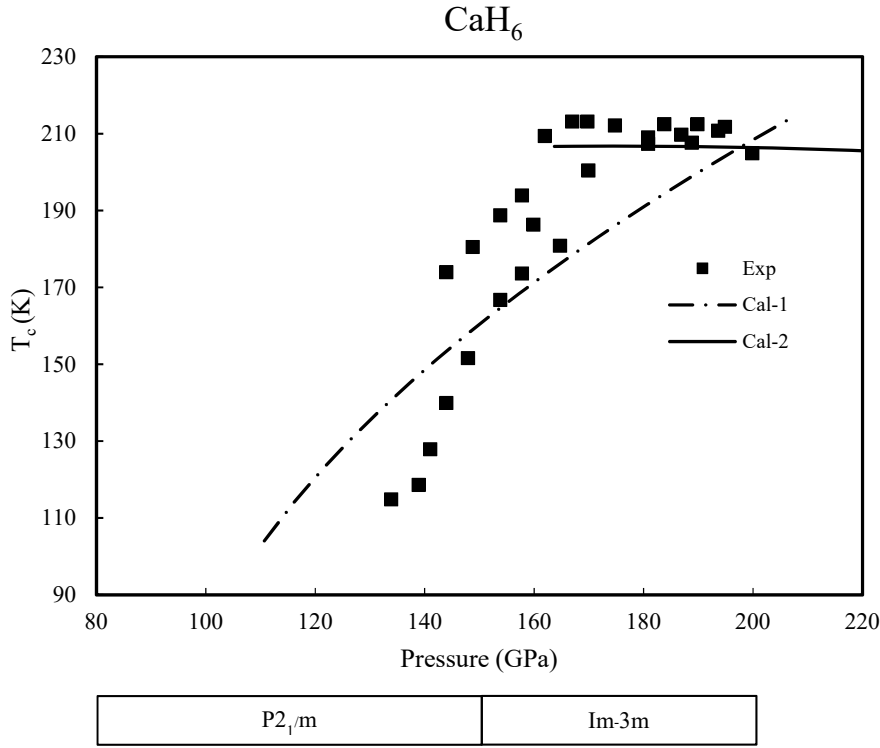

Figure.4. The critical temperature of $CaH_6$ superconductor was calculated (solid and dashed line) and measured experimentally (solid square) [9,10].

## 4. Conclusion

The critical temperatures under pressure of $H_3S$, $LaH_{10}$, $CaH_6$ and $Tl_2Ba_2CaCu_2O_{8,d}$ are investigated while performing under the constraint of weak coupling. The superconducting mechanism takes into account both the electron-phonon interaction and the Coulomb interaction. The equation of the critical temperature is calculated as a function of the unit cell volume fraction under high pressure. In order to determine the relationship between fraction volume and pressure, the Birch-Murnaghan equation of state is applied. Using this equation, we can investigate the relationship between superconductor's critical temperature and pressure. Cuprate superconductor and hydride superconductor are two types of superconductors that we would like to use. The phase transition in cuprate superconductors is caused by changes in the crystal structure, however the substance will continue to remain in the solid state even as the pressure increases. The phase transition of the hydride superconductor changes under high pressure; specifically, it goes from the gas phase to the solid phase during the process of increasing pressure. Since the fraction volumes of cuprate superconductors and hydride superconductors should be close to 1, we can impose constraints by determining the pressure- and volume-dependent factors. In cuprate superconductors, the experimental data for $Tl_2Ba_2CaCu_2O_{8,d}$ and our calculation are in good agreement. The superconducting hydride compounds $H_3S$, $LaH_{10}$, $CaH_6$ are investigated. There are separate lower and upper regions. These regions can be described by their parameters, and they can be well-fitted. The maximal critical temperature is predicted to be 112 K at 7 GPa, 197 K at 140 GPa, 252 K at 143 GPa, and 207 K at 174 GPa for the superconductors $Tl_2Ba_2CaCu_2O_{8,d}$, $H_3S$, $LaH_{10}$, $CaH_6$.